\documentclass{article}
\usepackage{spconf,amsmath,graphicx,hyperref}
\usepackage{booktabs,float}

\hypersetup{
    colorlinks=true,    
    linkcolor=green,     
    urlcolor=green,      
    citecolor=blue,
}

\title{Improving Anomalous Sound Detection via Low-Rank Adaptation Fine-Tuning of Pre-Trained Audio Models}
\name{Xinhu Zheng$^{1}$, Anbai Jiang$^{1}$, Bing Han$^{2}$, Yanmin Qian$^{2}$*, Pingyi Fan$^{1}$, Jia Liu$^{1}$, Wei-Qiang Zhang$^{1}$*\thanks{*Corresponding author} \thanks{This work was supported by the National Natural Science Foundation of China under Grant No. 62276153, and by a grant from the Guoqiang Institute, Tsinghua University.}}
\address{
$^1$ Department of Electronic Engineering, Tsinghua University, Beijing, China \\
$^2$ Department of Computer Science and Engineering, Shanghai Jiao Tong University, Shanghai, China\\
\tt wqzhang@tsinghua.edu.cn}

\begin{document}
%
\maketitle
\begin{abstract}
Anomalous Sound Detection (ASD) has gained significant interest through the application of various Artificial Intelligence (AI) technologies in industrial settings. Though possessing great potential, ASD systems can hardly be readily deployed in real production sites due to the generalization problem, which is primarily caused by the difficulty of data collection and the complexity of environmental factors. This paper introduces a robust ASD model that leverages audio pre-trained models. Specifically, we fine-tune these models using machine operation data, employing SpecAug as a data augmentation strategy. Additionally, we investigate the impact of utilizing Low-Rank Adaptation (LoRA) tuning instead of full fine-tuning to address the problem of limited data for fine-tuning. Our experiments on the DCASE2023 Task 2 dataset establish a new benchmark of 77.75\% on the evaluation set, with a significant improvement of 6.48\% compared with previous state-of-the-art (SOTA) models, including top-tier traditional convolutional networks and speech pre-trained models, which demonstrates the effectiveness of audio pre-trained models with LoRA tuning. Ablation studies are also conducted to showcase the efficacy of the proposed scheme.
\end{abstract}
\begin{keywords}
Anomalous sound detection, DCASE Challenge, pre-trained models, LoRA
\end{keywords}
\section{Introduction}
\label{sec:intro}

In the realm of industrial automation, the ability to detect anomalous sounds is crucial for maintaining operational integrity and preventing potential failures. The complexity of this task arises from the need to distinguish between normal operational noise and genuine anomalies, a process that requires sophisticated algorithms capable of learning from diverse acoustic patterns. In actual production environments, the diversity of equipment types, the complexity of the surroundings, and the presence of domain shift issues in sound data make it challenging to develop systems that can accurately identify and classify abnormal sounds across different devices and environments.

Anomalous sound detection tasks are distinct from other audio or speech tasks in several key ways~\cite{back, dcase23, JiangAB2023_ICASSP}. While tasks like speech recognition, speaker identification, and acoustic scene classification focus on identifying and categorizing sounds based on their content or context, ASD tasks are concerned with identifying sounds that deviate from a normative baseline. In ASD, the system must discern between normal operational sounds and those that are anomalous, which could indicate potential issues or malfunctions. This requires the system to have a deep understanding of what constitutes normalcy within a specific context, which is not always explicitly labeled in the training data. In contrast, other audio tasks often have more clearly defined categories and labeled datasets for training purposes. This makes ASD a unique and challenging area of research within the broader field of audio signal processing.

\begin{figure*}[t]
  \centering  \includegraphics[width=0.89\textwidth]{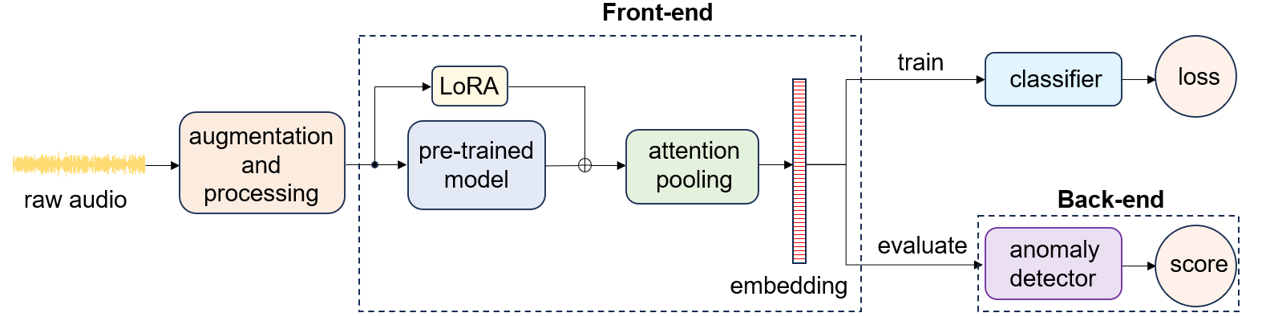}
  \label{fig:system}
  \caption{The overview of our system}
\end{figure*}

The field of anomalous sound detection has seen a surge of innovative research aimed at improving the accuracy and reliability of detecting abnormal sounds in various environments. Recent studies have explored a range of methodologies. In terms of unsupervised learning, traditional approaches have utilized statistical models and machine learning algorithms~\cite{wang2021unsupervised,jie2023anomalous}, with recent advancements incorporating deep learning techniques such as Autoencoders (AE) and Convolutional Neural Networks (CNN) for feature extraction and anomaly scoring~\cite{AE}. These methods often rely on the reconstruction error of AEs or the discriminative power of CNNs to differentiate between normal and anomalous sounds.

On the other hand, self-supervised techniques show great potential in recent works~\cite{GuanHEU2022, Giri2020, LiuCQUPT2022}. These techniques often involve data augmentation and pretext tasks that enable the model to learn from the data's inherent structure. For instance, the top-ranking team of DCASE 2020 Challenge~\cite{Giri2020} introduced an unsupervised anomalous sound detection method based on MobileNetV2 that leverages self-supervised classification and a group-masked autoencoder for density estimation. 

Recently, the application of pre-trained models in ASD has shown promising results~\cite{han2024exploring,jiang2024anopatch} in overcoming challenges such as limited labeled data and domain shifts. Systems in~\cite{LvHUAKONG2023} utilized pre-trained speech models to extract embeddings for ASD tasks, achieving 2nd place in the DCASE 2023 Challenge Task 2, showing the efficacy of leveraging pre-trained models in this domain.

The remarkable advancements in transformer-based pre-trained speech models have significantly impacted speech processing and related downstream tasks. Building upon this, researchers have developed pre-trained models tailored for audio processing, such as CED~\cite{ced} and BEATs~\cite{beats}, demonstrating promising performance in audio classification tasks and holding promise for various downstream applications.

Motivated by these efforts, we introduce a novel system based on self-supervised learning and pre-trained models. On the one hand, two prominent pre-trained speech models are utilized, namely Wav2Vec2~\cite{wav2vec2} and the encoder component of Qwen-Audio~\cite{qwen}. These models are distinguished by their training on vast speech datasets, which supports their ability to generalize across machine types and environments. Additionally, two audio-specific pre-trained models are employed, namely BEATs~\cite{beats} and CED~\cite{ced}, selected for their relevance to ASD tasks. Further, we employed Low-Rank Adaptation (LoRA)~\cite{lora} to fine-tune the models that exhibited superior performance. Our objective was to enhance their efficacy by preserving the initial features acquired during training and reducing the effects of overfitting. Our contribution can be summarized as follows:

\begin{itemize}
    \item We explore pre-trained audio models for anomalous sound detection of machines and compare them with widely used pre-trained speech models.
    \item We are the first to employ Low-Rank Adaptation for pre-trained audio models in ASD tasks. 
    \item The proposed method achieves SOTA on the DCASE 2023 Task2 dataset.
\end{itemize}

\section{Method}
\label{sec:format}

\subsection{System pipeline}
\label{ssec:system}
The system architecture is inspired by the 3rd ranking team~\cite{JiangTHUEE2023} of the DCASE 2023 Task 2 challenge, which is composed of two main components, namely the front-end and the back-end, as shown in Fig. \ref{fig:system}. The front-end is responsible for processing the original audio signals. It includes several pre-processing steps such as segmentation, windowing, enhancement, and converting to log-mel spectrogram. The processed audio is then fed into an audio model to extract semantic audio embeddings. During training, the audio embedding is further processed by a fully connected layer to perform classification tasks.

In the anomaly detection phase, the audio embedding is passed to the back-end, which consists of an anomaly detector. This detector processes the input audio embedding to determine whether it is anomalous and outputs an anomaly score. For practical applications of anomaly detection systems, an anomaly score thresholding approach is used to make judgments. In this paper, to evaluate the general performance, the area under the curve (AUC) and partial-AUC (pAUC) scores are then calculated as performance metrics.

\subsection{Data augmentation}
\label{ssec:data_aug}
In this work, we mainly utilize SpecAug~\cite{specaugment} as the data augmentation method. SpecAug operates directly on the feature inputs of a neural network, specifically the filter bank coefficients. It employs a straightforward augmentation policy that includes warping the features, masking blocks of frequency channels, and masking blocks of time steps. By manipulating these aspects of the audio signal, SpecAug effectively simulates various acoustic conditions, thereby enriching the diversity of the training dataset. However, only frequency masking and time masking are applied to the machine audio, while time warping is not applied due to the degradation of detection performance. Since Wav2Vec2 uses raw audio as input, SpecAug is not applied for this model.

\subsection{Pre-trained models}
\label{ssec:pretrain}
In this section, we will provide an overview of the pre-trained models employed in our system.

\textbf{Wav2Vec2}~\cite{wav2vec2} is a model for speech recognition tasks that employs self-supervised learning to extract speech representations from large amounts of unlabeled audio data. This allows for efficient semi-supervised training, where the model is first pre-trained on unlabeled data and then fine-tuned on a smaller labeled dataset. Remarkably, Wav2Vec2 can achieve state-of-the-art performance with as little as one hour of labeled data, demonstrating its ability to learn from limited labeled data and its potential for practical applications in ASD systems.

\textbf{Qwen-Audio}~\cite{qwen} is a multimodal large audio language model developed by Alibaba Cloud, which extends the Qwen series. The encoder component of Qwen-Audio is a pivotal element of its architecture, utilizing the Whisper-large-v2 model for initialization. It is designed to process a diverse array of audio inputs, including human speech, natural sounds, music, and songs. It is trained in a multi-task training framework The model achieves impressive performance across diverse benchmark tasks without requiring any task-specific fine-tuning.

\textbf{BEATs}~\cite{beats} introduces a novel approach to audio pre-training by utilizing discrete label prediction loss, which aligns with human auditory perception by focusing on high-level audio semantics rather than low-level time-frequency details. This method allows for the generation of discrete labels with rich audio semantics. This method allows for the generation of discrete labels with rich audio semantics. BEATs achieves this without the need for extensive training data or complex model parameters, setting new records such as a mean average precision (mAP) of 50.6\% on AudioSet-2M for audio-only models and 98.1\% accuracy on ESC-50.

\textbf{CED}~\cite{ced} introduces a novel approach to audio tagging by combining augmentation and knowledge distillation (KD) techniques. CED  utilizes the Vision Transformer (ViT) as its foundational model for audio tagging tasks. The ViT’s ability to handle variable-length inputs and its scalability makes it an effective choice for the CED framework, which aims to improve audio tagging performance through consistent ensemble distillation.

\subsection{Fine-tuning methods}
\label{ssec:finetune}
Two distinct fine-tuning strategies are employed to optimize the model’s performance. Firstly, we conducted full model fine-tuning. Secondly, we implemented Low-Rank Adaptation (LoRA) fine-tuning, a technique that introduces low-rank matrices to adapt the model’s weights without altering the original architecture. This method is particularly advantageous as it significantly reduces the computational resources required for fine-tuning, making it an efficient alternative to full model fine-tuning. More importantly, LoRA’s support for incremental learning enables the model to adapt to new tasks or data without complete retraining. By focusing on low-rank updates, LoRA helps preserve the knowledge captured during pre-training, maintaining performance on downstream tasks.

\subsection{Loss function and classification}
\label{ssec:loss}
Given the unpredictable and diverse nature of industrial equipment failures, the DCASE dataset~\cite{dataset1, dataset2} has not provided labels indicating whether devices are operating normally or not, nor has it supplied labeled anomalous data for model fine-tuning. Since our front-end model is a classifier, an appropriate proxy task must be identified. The DCASE dataset includes device type and attribute labels, with attributes encompassing specific device identifiers (A, B, C, D) and operational states (e.g., speed, gear position). These labels are utilized for the classification tasks, helping the models to learn the hidden features of the audio.

The choice of loss function is crucial during training. This paper employs the ArcFace loss~\cite{deng2019arcface}, an initially face recognition-focused loss function. The mathematical representation of ArcFace Loss is given by:

\begin{equation}
L_{\mathrm{ArcFace}} = -\frac{1}{N} \sum_{i=1}^{N} \log \frac{e^{s(\cos(\theta_{y_i} + m))}}{F}
\end{equation}
where $F=e^{s(\cos(\theta_{y_i} + m))} + \sum_{j=1, j\neq y_i}^{n} e^{s\cos(\theta_j)}$, $\theta_{y_i}$ is the angle between the feature vector $x_i$ and the weight vector $W_{y_i}$ of the ground-truth class $y_i$, and ${m}$ is the margin penalty.

In sound anomaly detection tasks, distinguishing between normal and abnormal sounds relies on the distance between audio features. ArcFace loss reinforces model discrimination by ensuring tight intra-class feature angles and dispersed inter-class feature angles, thus amplifying the difference between normal and abnormal samples and aiding in their effective separation.

\subsection{Back-end and anomaly detection}
\label{ssec:ano_detect}
In the back end of the system, we implement the K-Nearest Neighbors (KNN) algorithm as our anomaly detector, a method commonly utilized in ASD tasks. We have configured the hyper-parameter k to a value of 1 and selected cosine distance as our chosen metric for measuring distances. Additionally, in alignment with the approach detailed in~\cite{baseline}, we have trained our detector on datasets from both the source and target domains. The minimum distance recorded by these detectors is then employed as the anomaly score for each item. The final anomaly score for one item can be given by:

\begin{equation}
S_{\mathrm{Anomaly}}=\min(d_{\mathrm{source}},d_{\mathrm{target}})
\end{equation}
where $d_{\mathrm{source}}$ is the distance given by the detector trained in the source domain, and $d_{\mathrm{target}}$ is the distance given by the detector trained in the target domain.

\section{Experiments}
\label{sec:exps}

\subsection{Datasets}
\label{ssec:datasets}
In this paper, the DCASE 2023 Task 2 dataset is leveraged for evaluating the proposed scheme and comparison with previous SOTA models. The dataset features audio recordings from 14 machine types for both training and evaluation. Each machine type provides 1000 clips of 10-second sounds for training, including 990 clips in the source domain and 10 clips in the target domain. Additionally, the validation data is completely separate from the evaluation data based on machine types, and no domain information is provided for either set. The dataset is more challenging than classic anomaly detection datasets due to domain shift and rare data availability.

\subsection{Implementation details}
\label{ssec:impl}
In the data processing phase, the input configurations differ across models. For Wav2Vec2 and Qwen-Audio, which process the original audio files, each input segment is set to 2 seconds in duration. For BEATs and CED, which utilize Mel-frequency cepstral coefficients (MFCCs) as input, each frame is defined by a window length of 25 milliseconds, a hop length of 10 milliseconds, and utilizes a Hamming window function. The MFCCs employed are 128-dimensional.

For the training process, the learning rate scheduler is set to Warmup, with an initial learning rate of 0.00005 and an ADAM optimizer. We set the batch size to 8 and trained the model for 30 epochs.

For the results, the evaluation metrics are in line with the official evaluation metrics of the DCASE Challenge. AUC in both source and target domain and p-AUC are calculated, with $p$ in the range [0,0.1]. The final score is the harmonic mean of the three scores above across all machine types, which is also the official score used for ranking.

\subsection{Results and analysis}
\label{sec:results}
We initially compared the outcomes of the selected pre-trained models, including Wav2Vec2, Qwen-Audio, BEATs, and CED. The findings are summarized in Table~\ref{models}. For our baseline system, we selected the top-performing team from the DCASE 2023 Challenge, which utilized a multi-branch network incorporating CNN and Transformer architectures. The results revealed that models pre-trained on Audio-Set consistently outperformed all other models. In contrast, models pre-trained on speech-related tasks exhibited inferior performance compared to the baseline model which trained directly on the DCASE dataset.

\begin{table}[ht]
\centering
\caption{Performance (\%) of fine-tuning the pre-trained models. The models pre-trained on audio tasks outperform the models trained on speech tasks.}
\begin{tabular}{cccc}
\toprule
Model & dev & eval & hmean \\
\midrule
Wav2Vec2 & 60.97 & 57.65 & 59.26  \\
Qwen\_Audio & 59.27 & 52.58 & 55.73  \\
BEATs & \textbf{65.48} & \textbf{74.42} & \textbf{69.66} \\
CED & 64.60 & 66.93 & 65.74  \\
\bottomrule
\label{models}
\end{tabular}
\end{table}

Next, we tried using LoRA fine-tuning instead of full fine-tuning on the best-performing model, BEATs. Following the guidance of the LoRA authors, we initially introduced LoRA parameters to the $q$ and $v$ matrices of the Transformer architecture and set the hyper-parameter $r$ to 4. Given that LoRA was initially developed for NLP tasks, and considering the distinct nature of audio-related tasks, we explored various values for $r$, including 8, 16, 32, 64, and 128. The results presented in Table~\ref{r} indicate that for the task addressed in this paper, the optimal value of $r$ is 64. This outcome can be attributed to the fact that tasks such as ASD or similar audio-related challenges necessitate a greater number of hidden states compared to NLP tasks. Furthermore, the disparity between ASD tasks and pre-trained tasks contributes to this phenomenon, necessitating additional parameters for model adjustment to ASD-specific tasks.

\begin{table}[ht]
\centering
\caption{Performance (\%) of different values of $r$. Using LoRA can see an increase in the final scores, especially on the evaluation set.}
\begin{tabular}{cccc}
\toprule
$r$ & dev & eval & hmean \\
\midrule
Full fine-tune & 65.48 & 74.42 & 69.66   \\
4 & 65.64  & 76.77 &  70.77 \\
8 & 65.38 & 76.94 &  70.69 \\
16 & 64.58 & 76.35 & 69.97 \\
32 & 64.86 & 76.65 & 70.26 \\
64 & \textbf{67.26} & \textbf{77.16} & \textbf{71.87} \\
128 & 65.86 & 75.61 & 70.40 \\
\bottomrule
\label{r}
\end{tabular}
\end{table}

\begin{table}[H]
\vspace{-2.0em}
\centering
\caption{Performance (\%) of adding LoRA parameters in different matrices. The $v$ matrix of the Transformer shows more importance through the experiments.}
\begin{tabular}{cccc}
\toprule
Matrices & dev & eval & hmean \\
\midrule
$k$ & 64.00 & 74.69 & 68.93 \\
$q$ & 64.25 & 75.63 & 69.48  \\
$v$ & 65.19 & 75.12 &69.80 \\
$k$, $v$ & 65.13 & 75.82 &70.07 \\
$k$, $q$ & 64.53 & 74.93 & 69.34 \\
$k$, $q$, $v$ & 65.22 & 75.26 &69.88 \\
$q$, $v$ & \textbf{67.26} & \textbf{77.16} & \textbf{71.87} \\
\bottomrule
\label{kqv}
\end{tabular}
\end{table}

\setcounter{table}{3}
\begin{table*}[ht]
\centering
\caption{Performance (\%) of different systems. Our proposed methods achieve the best AUC score on most machine types, with a significant improvement of 6.48\% compared with previous SOTA models }
\begin{tabular}{ccccccccc}
\toprule
System & Bandsaw & Grinder & Shaker & ToyDrone & ToyNscale & ToyTank & Vacuum & hmean \\
\midrule
Jie et al.~\cite{jie2023anomalous} & 60.97 & 65.18 & 63.50 & 55.17 & 84.92 & 60.72 & 92.27 &  66.97  \\
Lv et al.~\cite{LvHUAKONG2023} & 55.47 & 64.76 & 70.98 & 52.89 & 71.90 & \textbf{70.73} & 91.48 & 66.39   \\
Jiang et al.~\cite{JiangTHUEE2023} & 60.36 & 63.90 & 60.12 & 53.75 & 82.70 & 61.96 & 91.10 & 65.63   \\
Zhang et al.~\cite{zhang2024dual} & - & - & - & - & - & - & - &  71.27 \\
Ours & \textbf{67.67} & \textbf{71.18} & \textbf{82.87} & \textbf{71.73} & \textbf{95.97} & 68.52 & \textbf{98.18} &  \textbf{77.75} \\
\bottomrule
\label{final}
\end{tabular}
\end{table*}

\setcounter{table}{4}
\begin{table}[ht]
\centering
\caption{Performance (\%) of adding LoRA parameters in different layers. The Transformer layers are divided into three parts. The layers closer to the output have a deeper impact on the results.}
\begin{tabular}{cccc}
\toprule
Layers & dev & eval & hmean \\
\midrule
1-4 & 63.37 & 71.61 & 67.24 \\
5-8 & 62.83 & 74.93 & 68.35 \\
9-12 & 64.23 & 75.87 & 69.57 \\
1-8 & 63.13 & 75.23 & 68.85 \\
1-4,9-12 & 63.99 & 76.05 & 69.50 \\
5-12& 64.09 & 77.38 & 70.11 \\
1-12 & \textbf{67.26} & \textbf{77.16} & \textbf{71.87} \\
\bottomrule
\label{layer}
\end{tabular}
\end{table}

\begin{table}[ht]
\vspace{-1em}
\centering
\caption{Performance (\%) of different adjustment strategies. Base refers to the best settings in the previous experiments. (1), (2), and (3) are the three strategies mentioned above. (3) achieves a new high on the evaluation set. }
\begin{tabular}{cccc}
\toprule
Adjustment & dev & eval & hmean \\
\midrule
Base & \textbf{67.26} & 77.16 & \textbf{71.87}  \\
$v$ 1.5 times(1) & 64.57 & 75.98 & 69.81 \\
latter half 1.5 times(2) & 65.30 & 76.88 & 70.62 \\
latter half $v$ 1.5 times(3) & 65.11 & \textbf{77.75} & 70.82 \\
\bottomrule
\label{fusion}
\end{tabular}
\end{table}

Besides this, we also conduct ablation studies. Since different layers and the various parameters within the same layer of a Transformer model will have distinct impacts on the results~\cite{lin2022survey}, this section continues to use the BEATs model as an example, by fine-tuning different layer numbers and different Transformer parameter matrices, to investigate how different layers and parameter matrices within the model affect the outcomes in ASD tasks. The following experiments first set some or all of the $k$, $q$, and $v$ parameters in each Transformer layer’s matrix to be trainable while freezing the rest of the parameters. Additionally, a series of experiments are conducted where LoRA parameters are introduced into different Transformer layers. The specific experimental results are detailed in the table below.

From the results in Table~\ref{kqv} and Table~\ref{layer}, it is evident that among the $k$, $q$, and $v$ parameter matrices of the Transformer, the $v$ matrix holds the greatest importance. Adjusting the k matrix tends to result in a relative performance decline. For the layers, those closer to the output tend to show more importance. However, adding LoRA parameters across all layers yields the best results. Combining these findings with previous LoRA tuning experiments, the final results achieve an optimal r value of approximately 64. Consequently, three adjustments are proposed for LoRA parameter dimensions: increasing the $v$ vector dimension by 1.5 times(1), increasing the dimension of the latter half of the layers by 1.5 times(2), and increasing the dimension of the latter half of the $v$ matrix by 1.5 times(3).

From the results in Table~\ref{fusion}, although these adjustment methods do not surpass previous methods in terms of average AUC scores on both training and testing sets, method (3) mentioned above results in a new high on the testing set. Therefore, by enhancing the number of trainable parameters for matrices that have a greater impact on results, it is possible to improve the model’s generalization ability to a certain extent. Also, we compare our results with the systems of the top 3 ranking teams in the DCASE 2023 Challenge and recent papers in Table~\ref{final}. Our method exceeds all the previous works on the evaluation set.

\section{Conclusion}
\label{sec:conclu}
This paper presents a robust ASD model that capitalizes on audio pre-trained models, which are fine-tuned with machine operation data and enhanced with SpecAugment for data augmentation. The research also explores the efficacy of Low-Rank Adaptation (LoRA) tuning as an alternative to full fine-tuning, addressing the challenge of limited data availability. Through ablation studies, we have optimized the integration of LoRA parameters. Our experimental results on the DCASE 2023 Task 2 dataset confirm that our methods surpass traditional convolutional networks and speech pre-trained models, validating the utility of audio pre-trained models and LoRA tuning. The achievement of a harmonic score of 77.75\% on the final evaluation dataset demonstrates the model’s superior performance. This work underscores the potential of audio pre-trained models and LoRA tuning in enhancing ASD’s generalization capabilities in complex industrial settings.

\newpage
\small
\bibliographystyle{IEEEbib}
\bibliography{refs}

\begin{thebibliography}{10}

\bibitem{back}
Yuki Tagawa, Rytis Maskeliūnas, and Robertas Damaševičius,
\newblock ``Acoustic anomaly detection of mechanical failures in noisy real-life factory environments,''
\newblock {\em Electronics}, vol. 10, no. 19, 2021,
\newblock Art. no. 2329.

\bibitem{dcase23}
Kota Dohi, Keisuke Imoto, Noboru Harada, Daisuke Niizumi, Yuma Koizumi, Tomoya Nishida, Harsh Purohit, Ryo Tanabe, Takashi Endo, and Yohei Kawaguchi,
\newblock ``Description and discussion on {DCASE} 2023 challenge task 2: First-shot unsupervised anomalous sound detection for machine condition monitoring,''
\newblock in {\em Proc. DCASE Workshop}, 2023, pp. 31--35.

\bibitem{JiangAB2023_ICASSP}
Anbai Jiang, Wei-Qiang Zhang, Yufeng Deng, Pingyi Fan, and Jia Liu,
\newblock ``Unsupervised anomaly detection and localization of machine audio: A {GAN}-based approach,''
\newblock in {\em Proc. ICASSP}, 2023.

\bibitem{wang2021unsupervised}
Yaoguang Wang, Yaohao Zheng, Yunxiang Zhang, Yongsheng Xie, Sen Xu, Ying Hu, and Liang He,
\newblock ``Unsupervised anomalous sound detection for machine condition monitoring using classification-based methods,''
\newblock {\em Applied Sciences}, vol. 11, no. 23, 2021,
\newblock Art. no. 11128.

\bibitem{jie2023anomalous}
Junjie Wang, Jiajun Wang, Shengbing Chen, Yong Sun, and Mengyuan Liu,
\newblock ``Anomalous sound detection based on self-supervised learning,''
\newblock Tech. {R}ep., DCASE 2023 Challenge, 2023.

\bibitem{AE}
Kevin Wilkinghoff,
\newblock ``Utilizing sub-cluster {AdaCos} for anomalous sound detection under domain shifted conditions,''
\newblock Tech. {R}ep., DCASE 2021 Challenge, 2021.

\bibitem{GuanHEU2022}
Feiyang Xiao, Youde Liu, Yuming Wei, Jian Guan, Qiaoxi Zhu, Tieran Zheng, and Jiqing Han,
\newblock ``The {DCASE2022} challenge task 2 system: Anomalous sound detection with self-supervised attribute classification and {GMM}-based clustering,''
\newblock Tech. {R}ep., DCASE 2022 Challenge, 2022.

\bibitem{Giri2020}
Ritwik Giri, Srikanth~V. Tenneti, Karim Helwani, Fangzhou Cheng, Umut Isik, and Arvindh Krishnaswamy,
\newblock ``Unsupervised anomalous sound detection using self-supervised classification and group masked autoencoder for density estimation,''
\newblock Tech. {R}ep., DCASE 2020 Challenge, 2020.

\bibitem{LiuCQUPT2022}
Ying Zeng, Hongqing Liu, Lihua Xu, Yi~Zhou, and Lu~Gan,
\newblock ``Robust anomaly sound detection framework for machine condition monitoring,''
\newblock Tech. {R}ep., DCASE 2022 Challenge, 2022.

\bibitem{han2024exploring}
Bing Han, Zhiqiang Lv, Anbai Jiang, Wen Huang, Zhengyang Chen, Yufeng Deng, Jiawei Ding, Cheng Lu, Wei-Qiang Zhang, Pingyi Fan, Jia Liu, and Yanmin Qian,
\newblock ``Exploring large scale pre-trained models for robust machine anomalous sound detection,''
\newblock in {\em Proc. ICASSP}, 2024, pp. 1326--1330.

\bibitem{jiang2024anopatch}
Anbai Jiang, Bing Han, Zhiqiang Lv, Yufeng Deng, Wei-Qiang Zhang, Xie Chen, Yanmin Qian, Jia Liu, and Pingyi Fan,
\newblock ``Anopatch: Towards better consistency in machine anomalous sound detection,''
\newblock in {\em Proc. Interspeech}, 2024, pp. 107--111.

\bibitem{LvHUAKONG2023}
Zhiqiang Lv, Bing Han, Zhengyang Chen, Yanmin Qian, Jiawei Ding, and Jia Liu,
\newblock ``Unsupervised anomalous detection based on unsupervised pretrained models,''
\newblock Tech. {R}ep., DCASE 2023 Challenge, 2023.

\bibitem{ced}
Heinrich Dinkel, Yongqing Wang, Zhiyong Yan, Junbo Zhang, and Yujun Wang,
\newblock ``{CED}: Consistent ensemble distillation for audio tagging,''
\newblock in {\em Proc. ICASSP}, 2024, pp. 291--295.

\bibitem{beats}
Sanyuan Chen, Yu~Wu, Chengyi Wang, Shujie Liu, Daniel Tompkins, Zhuo Chen, and Furu Wei,
\newblock ``{BEATs}: Audio pre-training with acoustic tokenizers,''
\newblock in {\em Proc. ICML}, 2022, pp. 5178--5193.

\bibitem{wav2vec2}
Alexei Baevski, Yuhao Zhou, Abdelrahman Mohamed, and Michael Auli,
\newblock ``wav2vec 2.0: A framework for self-supervised learning of speech representations,''
\newblock {\em Advances in Neural Information Processing Systems}, vol. 33, pp. 12449--12460, 2020.

\bibitem{qwen}
Yunfei Chu, Jin Xu, Xiaohuan Zhou, Qian Yang, Shiliang Zhang, Zhijie Yan, Chang Zhou, and Jingren Zhou,
\newblock ``{Qwen-Audio}: Advancing universal audio understanding via unified large-scale audio-language models,''
\newblock {\em arXiv preprint arXiv:2311.07919}, 2023.

\bibitem{lora}
Edward~J. Hu, Yelong Shen, Phillip Wallis, Zeyuan Allen-Zhu, Yuanzhi Li, Shean Wang, Lu~Wang, and Weizhu Chen,
\newblock ``{LoRA}: Low-rank adaptation of large language models,''
\newblock in {\em Proc. ICLR}, 2022.

\bibitem{JiangTHUEE2023}
Anbai Jiang, Qijun Hou, Jia Liu, Pingyi Fan, Jitao Ma, Cheng Lu, Yuanzhi Zhai, Yufeng Deng, and Wei-Qiang Zhang,
\newblock ``{THUEE} system for first-shot unsupervised anomalous sound detection for machine condition monitoring,''
\newblock Tech. {R}ep., DCASE 2023 Challenge, 2023.

\bibitem{specaugment}
Daniel~S Park, William Chan, Yu~Zhang, Chung-Cheng Chiu, Barret Zoph, Ekin~D Cubuk, and Quoc~V Le,
\newblock ``{SpecAugment}: A simple data augmentation method for automatic speech recognition,''
\newblock {\em Proc. Interspeech}, pp. 2613--2617, 2019.

\bibitem{dataset1}
Noboru Harada, Daisuke Niizumi, Daiki Takeuchi, Yasunori Ohishi, Masahiro Yasuda, and Shoichiro Saito,
\newblock ``{ToyADMOS2}: Another dataset of miniature-machine operating sounds for anomalous sound detection under domain shift conditions,''
\newblock in {\em Proc. DCASE Workshop}, 2021, pp. 1--5.

\bibitem{dataset2}
Kota Dohi, Tomoya Nishida, Harsh Purohit, Ryo Tanabe, Takashi Endo, Masaaki Yamamoto, Yuki Nikaido, and Yohei Kawaguchi,
\newblock ``{MIMII DG}: Sound dataset for malfunctioning industrial machine investigation and inspection for domain generalization task,''
\newblock in {\em Proc. DCASE Workshop}, 2022.

\bibitem{deng2019arcface}
Jiankang Deng, Jia Guo, Niannan Xue, and Stefanos Zafeiriou,
\newblock ``{ArcFace}: Additive angular margin loss for deep face recognition,''
\newblock in {\em Proc. CVPR}, 2019, pp. 4690--4699.

\bibitem{baseline}
Noboru Harada, Daisuke Niizumi, Daiki Takeuchi, Yasunori Ohishi, and Masahiro Yasuda,
\newblock ``First-shot anomaly detection for machine condition monitoring: A domain generalization baseline,''
\newblock {\em In arXiv e-prints: 2303.00455}, 2023.

\bibitem{zhang2024dual}
Yucong Zhang, Juan Liu, Yao Tian, Haifeng Liu, and Ming Li,
\newblock ``A dual-path framework with frequency-and-time excited network for anomalous sound detection,''
\newblock in {\em Proc. ICASSP}, 2024, pp. 1266--1270.

\bibitem{lin2022survey}
Tianyang Lin, Yuxin Wang, Xiangyang Liu, and Xipeng Qiu,
\newblock ``A survey of transformers,''
\newblock {\em AI Open}, vol. 3, pp. 111--132, 2022.

\end{thebibliography}
\end{document}